\documentclass[twocolumn,amsfonts]{revtex4-1}
\usepackage{graphicx}
\usepackage{amsmath}
\usepackage{lipsum}
\usepackage{stfloats}
\usepackage{bm}
\usepackage{float}
\usepackage{CJK}

\begin{document}

\title{Universal scaling behavior of resonant absorption}

\author{D. J. Yu }
\email{djyu79@gmail.com}
\affiliation{School of Space Research, Kyung Hee Uiversity\\ 1732, Deogyeong-daero, Yongin, Gyeonggi, 17104, Korea}

\author{T. \surname{Van Doorsselaere}}
\affiliation{Centre for mathematical Plasma Astrophysics, Department of Mathematics, KU Leuven\\ Celestijnenlaan
200B bus 2400, B-3001 Leuven, Belgium}

\begin{abstract}
Mode conversion and resonant absorption are crucial mechanisms for wave transport and absorption. Scaling behavior of mode conversion or resonant absorption is well-known for electromagnetic and MHD waves in planar geometry. Our recent study showed that such a scaling behavior of resonant absorption could also exist for coronal loop oscillations with cylindrical geometry, but it was only tested for one density profile.
Here we generalise our previous study on the scaling behavior of resonant absorption by considering multiple density profiles.
Applying an invariant imbedding method to the ideal MHD wave equations, we show that the scaling behavior also exists for these density models. We thus generalise our earlier results and show that such a universal scaling exists in cylindrical geometry, too. Given these results and the earlier results in planar geometry, we formulate a hypothesis that a universal scaling behavior exists regardless of the type of mode conversion or resonant absorption.
\end{abstract}

\maketitle

Mode conversion and resonant absorption can often occur in inhomogeneous plasmas such that two different waves are coupled with an exchange of energy. This mechanism has been considered as crucial for plasma heating and transport~\citep[][]{Chen1974,Ionson1978,Kivelson1986,Davila1987,Poedts1990,Steinolfson1993,Ofman1995,Nakariakov1999,Johnson2001,EHKim2007,EHKim2008,Okamoto2015}.
The previous literature is mainly concerned with mode conversion between electromagnetic waves and electrostatic waves or between magnetoacoustic waves and Alfv\'{e}n waves~\citep[e.g.,][]{Kivelson1986,Yu2016b}.
From theoretical and
numerical studies, it is well established that a scaling behavior exists for the mode conversion coefficient in a planar geometry
~\citep[][]{Forslund1975,Kivelson1986,Hinkel1989,Mjolhus1990,Hinkel1991,Hinkel1992,Hinkel1993,Willes2001,Kim_Lee2005,Kim_Lee2006,EHKim2008,Yu2013b,Schleyer2014}. The scaling parameter has in general a certain relation with the density gradient, wave vector, and wave frequency~\citep[see, e.g.,][]{Willes2001,Kim_Lee2005,Kim_Lee2006,EHKim2008,Schleyer2014}. The mode conversion coefficient, which denotes the ratio of energy transformation from one mode to another mode, has a maximum value of about 0.5 for the linear or parabolic density profile. When the density profile is more complex, the scaling behavior is broken and the conversion coefficient can vary greatly~\citep[][]{Willes2003,Yu2010,Yu2013a,Yu2016a}. While this scaling behavior is well-known in planar geometry, its behavior in cylindrical geometry was unknown until recently. In our recent study, we considered kink modes in coronal loops, where the coronal loop was assumed as a infinitely long, straight, and axisymmetric cylinder with a radial inhomogeneity. We found that resonant absorption of the kink modes in the Alfv\'{e}n resonance also show a certain scaling behavior. We called this behavior \textit{sub-universal} therein. However, we only considered a sinusoidal density profile in the transitional layer~\citep{Yu2016b}. Given that it was studied only for a particular density profile, it was unclear whether this behavior would be generally valid, as it is for planar geometry. This motivates the question if the results from our previous paper can be generalised to other density profiles. If they can be generalised, it would imply that the scaling behavior is universal in inhomogeneous plasma, regardless of the geometry, density profile, or mode conversion.

In this work, to answer this question if the scaling law also occurs for other density profiles, we extend our previous study on resonant absorption by comparing linear, sinusoidal, and parabolic density profiles. Adding to our previous results on the sinusoidal profile, we show that this kind of scaling behavior exists also for linear and parabolic density profiles. Thus, we show that the scaling behavior also exists in cylindrical geometry, for monotonic density profiles.  Based on these results, we speculate that a scaling feature of mode conversion (or resonant absorption) can generally occur regardless of the geometry, by synthesizing the newly-found scaling behaviors in cylindrical geometry with the previous ones in planar geometry.

The governing equation for the wave amplitude that describes the mode conversion usually contains a singularity in it, which makes it difficult to get exact solutions. The invariant imbedding method (IIM) transforms the concerned differential equations with boundary conditions into differential equations with initial conditions. It is an efficient and powerful tool for this kind of problems when one considers a one-dimensional inhomogeneity in the wave equations~\citep[][]{Babkin1980,Doucot1987,Rammal1987,Klyatskin1994,Kim1998,Kim_etal2001,Kim_Lee2005,Klyatskin2005}.
We apply the recently used invariant imbedding method~\citep[see][and references therein]{Yu2016b} to resonant absorption of externally-driven standing kink modes in the Alfv\'{e}n continuum.

From the ideal MHD equations, the wave equation for the perturbed total pressure $P$ under coronal conditions is written as,
by assuming a dependence $\exp[i(m\phi+k_zz-\omega t)]$ in cylindrical coordinates,~\citep[e.g.,][]{Soler2013,Yu2016b}
\begingroup
\small
\begin{eqnarray}
\frac{d^2 P}{d r^2}+\bigg[\frac{1}{r}-&&\frac{\omega^2}{\rho_0(\omega^2-\omega_A^2)}\frac{d\rho_0}{dr}\bigg]\frac{d P}{d r}+\bigg[\frac{\omega^2}{v_A^2}-k_z^2-\frac{m^2}{r^2}\bigg]P=0,\nonumber\\\label{eq:1}
\end{eqnarray}\endgroup
where $P\approx\mathbf{B}\cdot\mathbf{b}/\mu_0=B_{0}b_z/\mu_0$, $b_z$ is the perturbed magnetic field, the background magnetic field is $\mathbf{B}=(0,0,B_{0})$, $\omega_A^2(r)=k_z^2v_A^2=k_z^2B_{0}^2/\rho_0(r)\mu_0$, $k_z$ is the longitudinal wave number, $m$ is the azimuthal wavenumber, and $\mu_0$ is the magnetic permeability.

\begin{figure*}[]\center
\includegraphics[width=.35\textwidth]{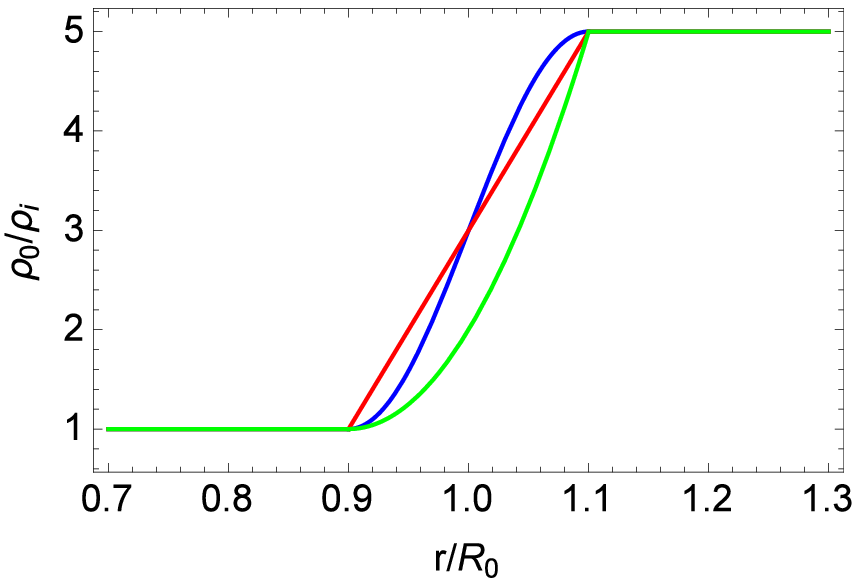}
\caption{\label{f1} Density profiles in the transitional layer when $l/R_0=0.2$ and $\Delta\rho(=\rho_e/\rho_i)=5$: linear (red), sinusoidal (blue), and parabolic (green) profiles.}
\end{figure*}

We assume that the density varies from $\rho_i$ inside the loop to $\rho_e$ outside the loop through the transitional layer of thickness $l$ ($\rho_t$):
\begingroup\small
\begin{eqnarray}
 \rho_0(r)= \left\{ \begin{array}{ll}
         \rho_i & \mbox{ if $r\leq R_0-l/2$ }\\
         \rho_t(r) & \mbox{ if $R_0-l/2<r<R_0+l/2$ }\\
         \rho_e & \mbox{ if  $r\geq R_0+l/2$}
        \end{array} \right.,\label{eq:2}
\end{eqnarray}\endgroup where $R_0$ represents the radius of the loop.
For the density $\rho_t(r)$ of the transitional layer we consider linear $(li)$, parabolic $(p)$, and sinusoidal $(s)$ profiles~\citep{Soler2013}:
\begingroup\small
\begin{eqnarray}
 \rho_{li}(r)&=& (\rho_e-\rho_i)(r-R_0)/{l}+(\rho_e+\rho_i)/{2}\label{eq:3},\\
 \rho_p(r)&=& \rho_i-(\rho_{i}-\rho_e)(r-R_0+l/2)^2/l^2\label{eq:4},\\
 \rho_s(r)&=& [(\rho_{e}+\rho_i)+(\rho_{e}-\rho_i)\sin(\pi(r-R_0)/l)] /2,\label{eq:5}
\end{eqnarray}\endgroup
 which are shown in Fig.~\ref{f1}.

For mode conversion and the related resonant absorption to occur, the wave frequency should reside in the Aflv\'{e}n continuum ($\omega_{Ae}<\omega<\omega_{Ai}$) of the loop. We consider that an external  wave is incident on a coronal loop with the frequency of the fundamental standing kink mode, $\omega_k=k_zB_0\sqrt{2/\mu_0(\rho_i+\rho_e)}$ with $k_z=\pi/L_0$, where $L_0$ is the loop length. Thus mode conversion occurs where $\omega_A=\omega_k$ or, in other words, $\rho(r)=(\rho_e+\rho_i)/2$. We consider the frequency dependence of mode conversion afterward. The wave outside the loop can be described with an incident wave and a scattered wave, which is
\begingroup\small
\begin{eqnarray}
P(r,\phi)&=&\sum_m a_m e^{i m\phi} \big\{J_m[k(r-R)+c]\nonumber\\
&&+H_m^{(1)}[k(r-R)+c]r_m(R)\big\},~~~~r> R,\label{eq:6}
\end{eqnarray}\endgroup
where $J_m(H_m^{(1)})$ is the Bessel (Hankel) function of the first kind, $r_m$ the scattering coefficient, $c$ a constant equivalent to $kR$, and $k(=\sqrt{(\omega^2/v_{A}^2)-k_z^2})$ the radial wave number for $r>R$. A constant $c$ is used both to apply the invariant imbedding method and for Eq.~(\ref{eq:6}) to be eigenfunctions for $r>R$~\citep[][]{Yu2016b}. The value $a_m$ depends on the form of the incident wave. For instance, when a plane wave is incident to the $x$ direction, $a_m=i^m$~\citep{Stratton2007}. Applying IIM to Eq.~(\ref{eq:1}) with Eq.~(\ref{eq:6}) and using the thickness $R$ as a new variable (imbedding parameter), we obtain for $r_m$~\citep{Yu2016b}
\begingroup\small
\begin{eqnarray}
\frac{dr_m(R)}{dR}&=&\frac{k}{H_m}\frac{\mu(R)}{\mu_1}[J_m'+{H_m^{(1)}}'r_m(R)]\label{eq:7}\\
&&+k\bigg[\frac{\mu(R)}{\mu_1}\frac{{H_m^{(1)}}'}{H_m^{(1)}}+\frac{1}{k R}\bigg]
\nonumber\\
&&\times\frac{[J'_m+{H_m^{(1)}}'r_m(R)][J_m+H_m^{(1)}r_m(R)]}{J'_mH_m^{(1)}-{H_m^{(1)}}'J_m}\nonumber\\
&&+k\bigg[\frac{\epsilon(R)}{\epsilon_1}
-\frac{\mu_1}{\mu(R)}\frac{m^2}{k^2 R^2}
\bigg]\frac{[J_m+H_m^{(1)}r_m(R)]^2}{J'_m H_m^{(1)}-{H_m^{(1)}}' J_m},\nonumber
\end{eqnarray}\endgroup
where $\mu(r)=\rho_0(r)(\omega^2-\omega_A^2(r))$, $\epsilon(r)=\mu_0/k_0^2B_0^2$, $k_0=\omega/c_0$, and $c_0$ is the speed of the electromagnetic wave in vacuum, and $\mu_1$ and $\epsilon_1$ are the values of $\mu$, $\epsilon$ for $r>R$, prime denotes $df(x)/dx$ for $f(x)$, $J_m=J_m(c)$, and $H_m^{(1)}=H_m^{(1)}(c)$. Eq.~(\ref{eq:7}) is a first order ordinary differential equation with $R$ as an integration variable. We integrate Eq.~(\ref{eq:7}) from $0$ to $R$ using initial conditions $r_m(0)=0$ to obtain $r_m(r=R)$~\citep{Yu2016b}, where the $1/kR$ term gives rise to a singularity at $R=0$ which we avoid by letting the values of parameters for $R<\delta$ equal those for $R=\delta\ll1$. This singularity is due to the cylindrical geometry we consider here. The value of $\delta$ needs to be sufficiently small to not affect the results. We also set $\omega=\omega_k+i\gamma$ to avoid the singularity at the resonance position ($\mu=0$) where $\gamma\ll1$, not affecting the results.
We define the mode conversion (absorption) coefficient $A$ as~\citep{Hanson2014,Yu2016b}
\begingroup\small
\begin{eqnarray}
A=-\sum_{m=\pm1}\textmd{Re}(r_m+|r_m|^2),\label{eq:8}
\end{eqnarray}\endgroup
where two kink modes ($m=\pm1$) are considered.

\begin{figure*}[]
\includegraphics[width=0.8\textwidth]{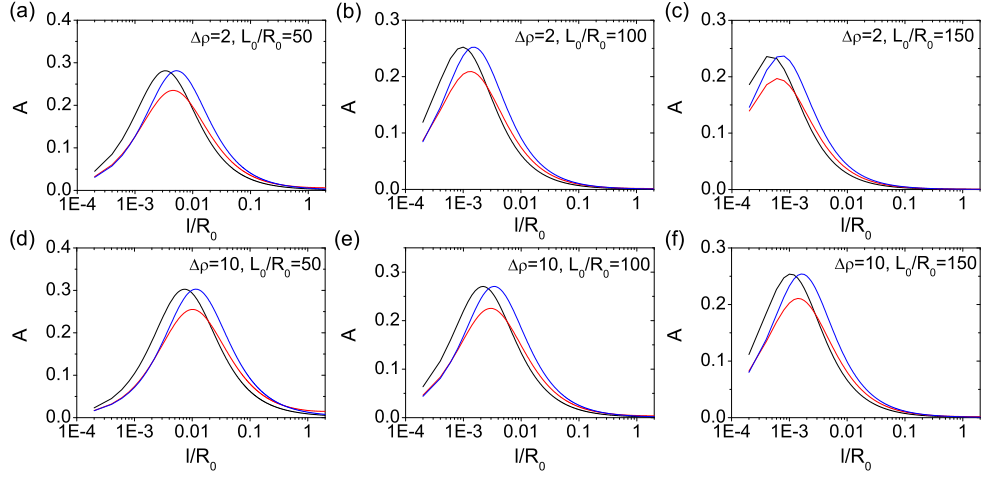}
\caption{\label{f2} Mode conversion coefficient ${A}$ for linear (black), parabolic (red), and sinusoidal (blue) density profiles as a function of $l/R_0$ for $\Delta\rho(=\rho_e/\rho_i)=$ 2 (top), 10 (bottom) and for $L/R_0=$ 50 (left), 100 (center), and 150 (right), where $\omega=\omega_k$, $k_z=\pi/L$, $R=3R_0$, $\rho_i=1.67353\times10^{-12}kg/m^3$, $B_0=10^{-3}T$, $R_0=2\times10^6m$, $\gamma=10^{-8}s^{-1}$, and $\delta=10^{-6}$. Mode conversion is strong when $l/R_0$ is small. }
\end{figure*}

In Fig.~\ref{f2}, we show the mode conversion coefficient $A$ versus $l/R_0$ for each density profile when $\Delta\rho(=\rho_e/\rho_i)=$ 2 and 10, and $L_0/R_0=$ 50, 100, and 150, while other fixed parameters are $\omega=\omega_k$, $k_z=\pi/L_0$, $R=3R_0$, $\rho_i=1.67353\times10^{-12}kg/m^3$, $B_0=10^{-3}T$, $R_0=2\times10^6m$, $\gamma=10^{-8}s^{-1}$, and $\delta=10^{-6}$.
We find that the mode conversion is strong when $l/R_0$ is small, depending on the loop length and density contrast. The peak positions shift towards a lower value of $l/R_0$ as $\Delta\rho$ decreases and $L_0/R_0$ increases. The figure shows that mode conversion for the parabolic profile is less effective than the other two density profiles, which is due to the difference of the resonance point.

\begin{figure*}[]
\includegraphics[width=0.8\textwidth]{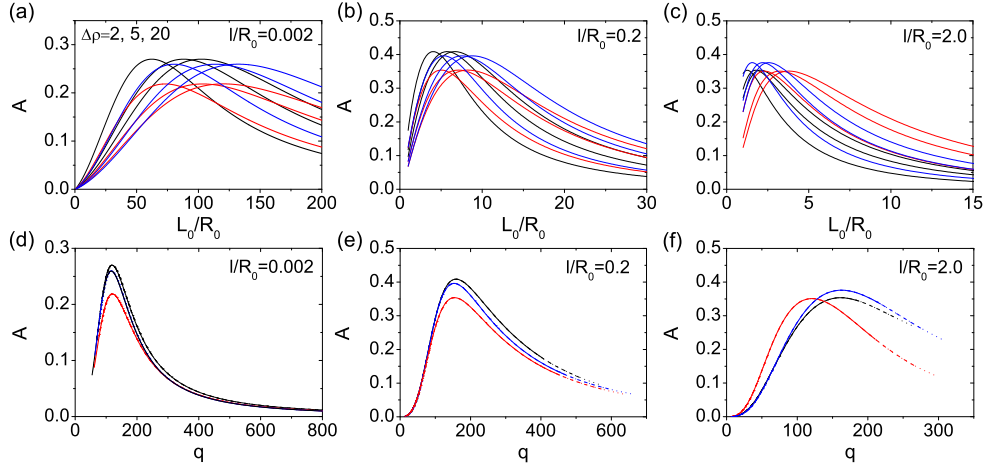}
\caption{\label{f3} Mode conversion coefficient ${A}$ for linear (black), parabolic (red), and sinusoidal (blue) density profiles as a function of $L_0/R_0$ (a)-(c) and of $q$ (d)-(f) for $l/R_0=$ 0.002 (left), 0.2 (center) and 2.0 (right), where $\Delta\rho=$ 2, 5, and 20. In (a)-(c), as $\Delta\rho$ increases the peak position of the curves shifts toward the higher values of $L_0/R_0$. As a function of $q$, $A$ reduces into a single curve for each density profiles, while the curves still depend on the value of $l/R_0$. The scaling parameter $q$ is defined as $[c_1(R_0/l)(\Delta\rho-1)/(\Delta\rho+1)]^{1/3}(k_zR_0)^{2/3}$ where $c_1$ corresponds to $\pi$, $2\sqrt{2}$, and $2$ for linear, parabolic, and sinusoidal profiles, respectively. The other parameters are the same as in Fig. 2. }
\end{figure*}

In Fig.~\ref{f3} (a)-(c), we plot the mode conversion coefficient $A$ versus $L_0/R_0$ for each density profile when $l/R_0=$ 0.002, 0.2 and 2.0, and $\Delta\rho=$ 2, 5, and 20. For each $l/R_0$, the peak value of $A$ for each density profile has the same value regardless of $\Delta\rho$, which draws us to a finding below. The mode conversion coefficient can be plotted as a single curve when expressed with a scaling factor for some simple density (or Alfv\'{e}n speed) profiles~\citep{Forslund1975,Kivelson1986,Hinkel1992,Hinkel1993,Kim_Lee2005}. Approximating the density profile near the resonance (mode conversion) position $r=R_*$ as a linear function $\rho_t(r)=a(r-R_*)+b$ ($a$ and $b$ are constants) and including $\omega=\omega_k+i\gamma$, the wave equation reduces to~\citep{Yu2016b}
\begingroup
\small
\begin{eqnarray}
\frac{d^2 P}{d \bar{r}^2}&+&\bigg(\frac{1}{\bar{r}}-\frac{1}{\bar{r}-q+i\bar{\gamma}}\bigg)\frac{d P}{d \bar{r}}+\bigg(\bar{r}-q+i\bar{\gamma}-\frac{m^2}{\bar{r}^2}\bigg)P=0,\nonumber\\\label{eq:9}
\end{eqnarray}\endgroup
where $\bar{r}=\kappa r$, $q=\kappa R_*$, $\bar{\gamma}=2(a/b)^{-2/3}{k_z}^{2/3}(\eta/\omega_k)$ and $\kappa=(a/b)^{1/3}(k_z)^{2/3}$. In the limit $\bar{\gamma}~(\gamma)\to 0$, this equation only depends on the value of $q$ for a given $m$~\citep{Kim_Lee2005,Kim_etal2008,Yu2016b}. Thus one can expect a single curve of the mode conversion coefficient in terms of $q$ for simple density profiles. For the linear, sinusoidal, and parabolic density profiles, we obtain that $q$ equals $[(\pi R_0/l)(\Delta\rho-1)/(\Delta\rho+1)]^{1/3}(k_zR_0)^{2/3}$, $[(2R_0/l)(\Delta\rho-1)/(\Delta\rho+1)]^{1/3}(k_zR_0)^{2/3}$, and $[(2\sqrt{2}R_0/l)(\Delta\rho-1)/(\Delta\rho+1)]^{1/3}(k_zR_0)^{2/3}$, respectively~\citep[cf.,][]{Yu2016b}. With those scaling factors, the mode conversion coefficient $A$ in Fig.~\ref{f3} (a)-(c) reduces to a single curve in Fig.~\ref{f3} (d)-(f) for each density profile. A difference from the results in planar geometry is that $A$ has some dependence on $l/R_0$ despite $q$ is a function of $l/R_0$. However, Fig.~\ref{f3} points out that the scaling behavior is a general property if the density profile can approximately be linearized at the resonance point, reinforcing and extending the results from planar geometry. It is straightforward to apply these results to the mode conversion of electromagnetic waves into electrostatic modes by using the mathematical analogy of the two governing wave equations~\citep[][]{Kim_Lee2005,Kivelson1986,Yu2016b}. Comparing the scaling behavior in planar and cylindrical geometries, we propose that a scaling behavior can exist regardless of the geometry for density configurations that are approximately linear around the resonant point.

\begin{figure*}[]
\includegraphics[width=0.8\textwidth]{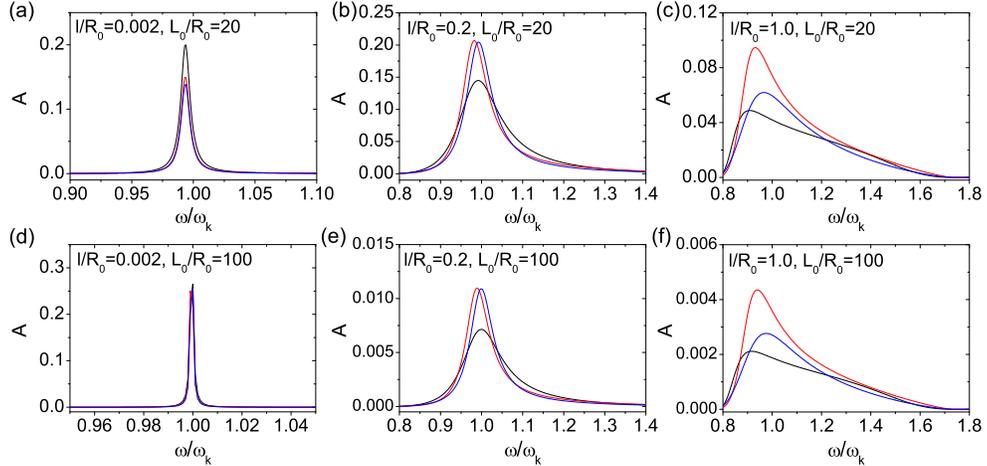}

\caption{\label{f4}  Mode conversion coefficient ${A}$ for linear (black), parabolic (red), and sinusoidal (blue) density profiles as a function of $\omega/\omega_k$ for $\Delta\rho$=5, $l/R_0$=0.002 (left), 0.2 (center) and 1.0 (right), and $L_0/R_0$=20 (top) and 100 (bottom). Mode conversion is strong when $\omega\approx\omega_k$ and $l/R_0$ is small. The relative relations of $A$ among the density profiles do not depend on the value of $L_0/R_0$. The other parameters are the same as in Fig. 2.}
\end{figure*}

In contrast with previous results, we here consider the frequency dependence of the mode conversion.
Fig.~\ref{f4} presents the dependence of the mode conversion on the wave frequency.
We draw ${A}$ versus $\omega/\omega_k$ for $\Delta\rho$=5, $l/R_0$=0.002, 0.2 and 2.0, and $L_0/R_0$=20 and 100. Mode conversion is strongest when $\omega\approx\omega_k$ and $l/R_0$ is small, but the peak position shifts downwards as $l/R_0$ increases. Comparing the top and bottom rows, the shape of the mode conversion coefficient is almost independent of $L_0/R_0$.

\label{sec4}
In this paper, we have generalised our earlier results on universal scaling in resonant absorption to systems with different density profiles. We have done this using the invariant embedding approach. We found, irrespective of the particular density profile, that mode conversion (resonant absorption) is efficient when the thickness of the transitional layers ($l/R_0$) is small and the loop length ($L_0/R_0$) is relatively large. From this dependence irrespective of the density profile (as long as it is approximately linear near the resonant point), we thus infer that a universal scaling behavior of mode conversion is found for coronal loops exhibiting kink oscillations. In this particular case, the scaling formulas is useful, because it could be possible to infer the wave energy absorption due to resonant absorption from the observed wave parameters~\citep{Arregui2018}. We expect that a similar scaling behavior is expected for resonant absorption of usual coronal loop oscillations for which the density contrast $\Delta\rho<1$. \par

The results of our study also have implications beyond the field of solar physics. Our results show that the universal scaling behaviour for mode conversion in resonant absorption extends to cylindrical geometry, adding to the previous knowledge on planar geometry. The universal scaling is thus truely universal, and happens irrespective of geometry. It is generally applicable to other forms of mode conversion as well. In particular, we can infer that the universal scaling also applies to mode conversion of electromagnetic waves into electrostatics waves in unmagnetized plasmas or in metamaterials~\citep{Ding2013,Kim_etal2008,Sun2015} in cylindrical (or spherical) geometry, because that is also described with the same mathematical analogy of the wave equations~\citep[][]{Yu2016b} as here. Furthermore, going beyond mode conversion of electromagnetic or magnetohydrodynamic waves, other kind of mode conversion like e.g. between gravitational waves and electromagnetic waves are also possible~\citep{Brodin1999,Marklund2000}. Our results allow us to postulate that, even for these exotic cases, a scaling behavior also exists. We can thus conclude the paper by formulating the hypothesis: ``a scaling behavior of the mode conversion coefficient is possible for monotonic density profiles regardless of the geometry and the type of mode conversion.''

\acknowledgments
This work was based on the presentation in 43rd European Physical Society (EPS) Conference on Plasma Physics (2016). D.J.Y. thanks the support from the BK21 plus program through the National Research Foundation (NRF) funded by the Ministry of Education of Korea. T.V.D. thanks the support from GOA-2015-014 (KU Leuven), and the European Research Council (ERC) under the European Union¡¯s Horizon 2020 research and innovation programme (grant agreement No. 724326).

\bibliographystyle{aipauth4-1}
\bibliography{mc_kink0_ref}




\end{document}